# Giant Piezospintronic Effect in a Noncollinear Antiferromagnetic Metal

Huixin Guo, Zexin Feng, Han Yan, Jiuzhao Liu, Jia Zhang, Xiaorong Zhou, Peixin Qin, Jialin Cai, Zhongming Zeng, Xin Zhang, Xiaoning Wang, Hongyu Chen, Haojiang Wu, Chengbao Jiang and Zhiqi Liu*

H. Guo, Z. Feng, H. Yan, X. Zhou, P. Qin, X. Zhang, X. Wang, H. Chen, H. Wu, Prof. C. Jiang, Prof. Z. Liu
School of Materials Science and Engineering, Beihang University, Beijing 100191, China
E-mail: zhiqi@buaa.edu.cn
J. Liu, Prof. J. Zhang
School of Physics and Wuhan National High Magnetic Field Center, Huazhong University of Science and Technology, Wuhan 430074, China
J. Cai, Prof. Z. Zeng
Key Laboratory of Multifunctional Nanomaterials and Smart Systems, Suzhou Institute of Nano-Tech and Nano-Bionics, Chinese Academy of Sciences, Suzhou 215123, China

**One of the main bottleneck issues for room-temperature antiferromagnetic spintronic devices is the small signal read-out owing to the limited anisotropic magnetoresistance in antiferromagnets. However, this could be overcome by either utilizing the Berry-curvature-induced anomalous Hall resistance in noncollinear antiferromagnets or establishing tunnel junction devices based on effective manipulation of antiferromagnetic spins. In this work, we demonstrate the giant piezoelectric strain control of the spin structure and the anomalous Hall resistance in a noncollinear antiferromagnetic metal - $D0_{19}$ hexagonal $Mn_3Ga$. Furthermore, we built tunnel junction devices with a diameter of 200 nm to amplify the maximum tunneling resistance ratio to more than 10% at room-temperature, which thus implies significant potential of noncollinear antiferromagnets for large signal-output and high-density antiferromagnetic spintronic device applications.**

Current spintronic devices are mainly based on ferromagnetic materials. Compared with ferromagnets, antiferromagnets have the insensitivity to external magnetic fields, ultrahigh spin dynamics frequency of THz which provides a prerequisite for picosecond operation of the spintronic devices and high packing density owing to the absence of fringing stray fields. Therefore, antiferromagnetic spintronic devices could have significant impacts on the fields of magnetic random access memories, artificial neural networks, THz memory-logic devices and THz sources and detectors.

The recent theoretical studies have found that the anomalous Hall effect (AHE) is intrinsically induced by Berry curvature, which is a pseudo magnetic field in momentum space and originates from the topological bands interactions of Bloch electrons. The anomalous Hall conductivity can be obtained through an integration over all occupied states below the Fermi energy.[1] Owing to the nonvanishing Berry curvature, noncollinear antiferromagnets with zero net magnetization can exhibit the AHE[2,3] such as in cubic $Mn_3Ir$, cubic $Mn_3Pt$, hexagonal $Mn_3Sn$ and $Mn_3Ge$. In turn, the anomalous Hall effect can behave as an electric probe to the spin structure in noncollinear antiferromagnets.

Motivated by some pioneering studies,[4-6] the field of antiferromagnetic spintronic has been substantially advanced.[7-14] Up to now, antiferromagnetic spintronic devices are mainly based on the anisotropic magnetoresistance effect in antiferromagnets,[5,6] which is typically on the order of 0.1% and is too small for real applications. To overcome this urgent shortcoming, the anomalous Hall resistance in noncollinear antiferromagnets and antiferromagnetic tunnel junction devices that reply on the effective control of antiferromagnetic spins and tunneling resistance could be promising for enhancing the spintronic signal output as we have recently proposed.[11]

In addition, although antiferromagnetic spins are challenging to be controlled by magnetic fields, the piezoelectric strain triggered by electric fields in ferroelectric materials could be an effective low-power approach to harnessing antiferromagnetic spintronic devices.[15-19] In this work, we report the important progress on electric-field-controlled anomalous Hall resistance and tunneling resistance based on a noncollinear antiferromagnetic metal $Mn_3Ga$. Similar to hexagonal $Mn_3Sn$,[20] $Mn_3Ga$ with a hexagonal $D0_{19}$ phase is a noncollinear antiferromagnet.[21,22] It has a weak ferromagnetic moment at temperatures lower than its Néel temperature (~470 K) due to the slightly canted triangular spin structure. The lattice parameters are $a$ = 5.404 and $c$ = 4.357 Å.[23] 50-nm-thick $Mn_3Ga$ thin films were grown on (100)-oriented ferroelectric $0.7PbMg_{1/3}Nb_{2/3}O_3$–$0.3PbTiO_3$ (PMN-PT) single-crystal substrates by sputtering at 300˚C. Considering the situation that the Pb in PMN-PT substrates is volatile at high temperatures and it would diffuse into the $Mn_3Ga$ film to form secondary alloys and subsequently lead to the degradation of surface roughness, 300˚C is an optimized growth temperature to ensure the crystallinity and small surface roughness of $Mn_3Ga$.

**Figure 1**a shows the cross-section transmission electron microscopy image of a $Mn_3Ga$/PMN-PT heterostructure. It reveals the absence of any interfaial second phase between the $Mn_3Ga$ film and the PMN-PT substrate. The x-ray diffraction data of the $Mn_3Ga$/PMN-PT heterostructure in Figure 1b indicates the strong (200) peak of the hexagonal $D0_{19}$ phase of $Mn_3Ga$. The magnetic-field dependent magnetic moment curve of the $Mn_3Ga$/PMN-PT sample measured at 300 K is presented in Figure 1c. It is dominated by the diamagnetic signal of the PMN-PT substate, which implies that the $Mn_3Ga$ thin film has a rather weak canted magnetic moment at room temperature, which is consistent with the expected antiferromagnetic nature of the $Mn_3Ga$ film.

To further explore the magnetic order of the $Mn_3Ga$ film, a 5-nm-thick $Co_{90}Fe_{10}$ (CoFe) thin film was grown on a $Mn_3Ga$/PMN-PT heterostruture with a 2-nm-thick capping layer Pt film via sputtering at room temperature. As shown in Figure 1d, a large exchange bias $\mu_0 H_{EB}$ of ~28.0 mT and a large coercivity field $\mu_0 H_C$ of 127.6 mT are obtained at 50 K for the CoFe/$Mn_3Ga$ heterostructure. The large increase of the CoFe coercivity field compared with a coercivity field of a single CoFe layer grown on MgO[24] and the appearance of the large exchange bias are evidence of strong exchange coupling of the $Mn_3Ga$/CoFe bilayer system, which thus proves the antiferromagnetic order of the $Mn_3Ga$ film.

Subsequently, we examined the AHE of the $Mn_3Ga$/PMN-PT heterostructure utilizing the Hall measurement geometry schematized in **Figure 2**a. The anomalous Hall resistance as a function of magnetic field presents obvious hysteresis loops at all the temperatures ranging from 300 K to 50 K (Figure 2b-d). As the temperature decreases, the Hall resistance at zero magnetic field increases from ~0.112 Ω to ~0.364 Ω and the coercivity field for switching the anomalous Hall resistance has a noticable enhancement from 93 mT to 667.6 mT. This is consistent with the Berry-curvature-induced AHE in a noncollinear antiferromagnet that has a negligible magnetic moment but a large anomalous Hall resistance. The switching of the anomalous Hall resistance by a magnetic field for noncollinear antiferromagnets relies on the magnetic switching of the canted moments. Therefore, it is similar to the magnetization switching in ferromagnets and the switching field increases with lowering temperature.

Next we analyzed the effect of the piezoelectric strain on the anomalous Hall resistance by applying a gate electric-field $E_G$ of 4 kV/cm perpendicularly to the PMN-PT substrate as the electrostatic modulation mechanism could hardly work for 50-nm-thick $Mn_3Ga$ metal films. The measurement geometry is illustrated in Figure 2e. To our surprise, the anomalous Hall resistance at $E_G$ = 4 kV/cm exhibits a giant change at all the temperatures (Figure 2f-h). Taking the data obtained at 50 K as an example, the zero-magnetic-field Hall resistance changes from ~0.364 Ω at $E_G$ = 0 kV/cm by more than one order of magnitude, to ~0.010 Ω at $E_G$ = 4 kV/cm. As the AHE in noncollinear antiferromagnets is a sensitive probe to their spin structures, the giant change of the AHE under the piezoelectric strain indicates the large variation of the spin structure.

The piezoelectric strain of a ferroelectric PMN-PT substrate highly relies on the non-180° ferroelastic ferroelectric polarization switching.[25] As shown in **Figure 3**a, there are 8 equivalent spontaneous polarization directions along <111> orientations in its unpoled state. After applying a positive electric field of 4 kV/cm to the PMN-PT substrate, the ferroelectric domains are switched upwards and an in-plane biaxial compressive piezoelectric strain is generated. The magnetic-field dependence of magnetization of the unpoled sample is shown in Figure 3b. In contrast, a weak ferromagnetic magnetization of ~6 emµ/cc is found for the Mn$_3$Ga thin film after poling the PMN-PT substrate by $E_G$ = 4 kV/cm at 300 K (Figure 3e). This suggests that the spin canting of the noncollinear spin structure in Mn$_3$Ga is increased under piezoelectric strain, which is plausible as similar to Mn$_3$Sn the hexagonal Mn$_3$Ga with inverse triangular spin structure has weak magnetic anisotropy[20,21] and thus its spin structure can be sensitive to strain-induced magnetoelastic energy.

To further explore the impact of strain on the magnetic properties of Mn$_3$Ga thin film, we measured the magnetic properties of the Pt/CoFe/Mn$_3$Ga/PMN-PT heterostructure before and after electric excitation by a positive electric field of 4 kV/cm. Compared with the unpoled sample at 50 K (Figure 3c), the exchange bias field $\mu_0 H_{EB}$ of the heterostructure is largely increased by ~66%, up to ~46.6 mT (Figure 3f) after electric poling. In addition, the coercivity field of the poled sample is enhanced by ~10% to 141.9 mT relative to the unpoled heterostructure. Considering that the magnetic properties of the thin CoFe film is not sensitive to piezoelectric strain,[26] the increase of the exchange bias and coercivity fields confirms the spin structure change of noncollinear antiferromagnet Mn$_3$Ga.

To quantify the in-plane piezoelectric strain, a strain gauge is attached to the PMN-PT substrate (**Figure 4**a) to measure the strain-electric field curve. The consequence of the $E_G$ dependent in-plane strain of the PMN-PT single-crystal substrate is illustrated in Figure 4b. As clearly seen, $E_G$ alters the in-plane strain up to ~0.1% between the unpoled state and the state at $E_G$ = 4 kV/cm because of 109° ferroelastic domain switching.[27,28]

According to the increased spin canting of Mn$_3$Ga under piezoelectric strain, Figure 4c shows a schematic scenario where the spins are rotated by piezoelectric strain. Furthermore, theoretical calculations were performed using the Quantum Espresso package[29] with the PBE-GGA (Perdew-Burke-Ernzerhof type of generalized-gradient-approximation) exchange correlation potential[30] and ultrasoft pseudopotential[31] generated from PSlib0.3.1. A Monkhorst-Pack *k*-point mesh of 16×16×20 and plane-wave cutoff 50 Ry are adopted for the self-consistent electronic structure calculations of bulk $Mn_3Ga$. The Fermi surface is plotted by using Xcrysden.[32] It was found that the density of states at the Fermi energy is very sensitive to the spin rotation angle in noncollinear $Mn_3Ga$ (Figure 4d). In addition, the Fermi surface could be largely modified by the spin structure change in $Mn_3Ga$ (Figure 4e). As the tunneling anisotropic magnetoresistance is highly dependent on spin-dependent density of states,[33,34] these results thus provide a rather attractive path to realizing antiferromagnetic tunnel junctions based on the piezoelectric-strain-modified spin structure and density of states in noncollinear antiferromagnet $Mn_3Ga$.

Therefore, we further fabricated nanoscale tunnel junction devices via standard industry-level tunnel junction fabrication procedures based on high-precision e-beam lithography technique. **Figure 5**a shows an optical microscope image of our magnetic tunneling junctions (MTJs) with a central pillar junction structure that has a diameter of 200 nm. The stacking structure is $Mn_3Ga$(50 nm)/MgO(2 nm)/Pt(10 nm)/Ti(20 nm)/Au(100 nm) and is shown in Figure 5b. The MgO layer is the tunneling barrier and the Ti and Au layers are used as top electrodes. A 50-nm-thick $SiO_2$ was deposited in the cross-area between the top electrodes and $Mn_3Ga$ thin film to separate them and also protect $Mn_3Ga$ from oxidation. Figure 5c depicts the measurement geometry of electric-field-controlled MTJ. A constant voltage bias of 50 mV was applied across the junction through the top and the bottom elctrodes. The gate electric field was applied between the bottom gate and the top electrode across the PMN-PT substrate. We then analyzed the influence of the gate electric field on the tunneling resistance of the MTJ. The resistance of a typical device is plotted as a function of $E_G$ in Figure 5d. It is found that the shape of the curves is asymmetrc bufferfly loops similar to asymmetric strain curve in (001)-oriented PMN-PT[26-28] with high stability.

The maximum electric-field-modulated resistance variation is ~10.7% for $E_G$ = -4 and +4 kV/cm. Two different resistance states, ~65699 Ω (~2064 Ω·μm$^2$) and ~59842 Ω (~1880 Ω·μm$^2$), are present at $E_G$ = 0 kV/cm. This corresponds to a nonvolatile tunneling resistance modulation ratio of ~9.8%. The peaks of the gate current ($I_G$) prove the polarization switching in the PMN-PT substrate (Figure 5e). In addition, electric pulses of $E_G$ = -3 kV/cm and $E_G$ = +3 kV/cm produce nonvolatile high-resistance and low-resistance states, respectively (Figure 5f). This thus demonstrates an excellent electric-field-controlled antiferromagnetic tunnel junction memory. It is worth emphasizing that the large room-temperature tunneling resistance modulation in our device may not fully come from the spintronic origin as the insulating barrier layer could be deformed by piezoelectric strain and would thus affect the tunneling process.

To summarize, we have successfully fabricated $D0_{19}$ hexagonal noncollinear antiferromagnetic Mn$_3$Ga thin films on ferroelectric PMN-PT substrates. The anomalous Hall resistance of noncollinear Mn$_3$Ga is largely modulated by piezoelectric strain, which suggests that the spin structure of noncollinear antiferromagnetic Mn$_3$Ga is rather susceptible to the piezoelectric strain. More importantly, based on the piezoelectric-strain-modified spin structure and density of state in noncollinear antiferromagnet Mn$_3$Ga, we have achieved an effective electric-field-controlled antiferromagnetic MTJ memory device. This work thus opens a brand new path to utilizing noncollinear antiferromagnets for large signal read-out, ultrafast, energy-efficient and high-density spintronic device applications.

**Experimental Section**

*Growth:* Mn$_3$Ga thin films were grown on 0.7PbMg$_{1/3}$Nb$_{2/3}$O$_3$–0.3PbTiO$_3$(PMN-PT) substrates using sputtering technique with a base pressure of 7.5×10$^{-9}$ Torr. During the depositing procedure, the sputtering power, the Ar pressure and the growth temperature were

d.c. 60 W, 3mTorr and 300 ˚C, respectively. As for the Pt/Co$_{90}$Fe$_{10}$/Mn$_3$Ga/PMN-PT heterostructure, a Pt film and a Co$_{90}$Fe$_{10}$ film were obtained by the d.c. sputtering system at room temperature at a Ar pressure of 3 mTorr. The sputtering power of the Pt thin film and the Co$_{90}$Fe$_{10}$ thin film were 30 W and 90 W, respectively. The growth rate of Pt was ~0.5 Å/s and Co$_{90}$Fe$_{10}$ was ~0.11 Å/s.

When it turns to the Mn$_3$Ga(50 nm)/MgO(2 nm)/Pt(10 nm)/Ti(20 nm)/Au(100 nm) magnetic tunnel junctions, the Mn$_3$Ga layers were grown in the same conditions as mentioned above. The MgO and the Pt layer were fabricated by a plused laser deposition system with a laser fluence of ~1.6 J/cm$^2$ and a repetition rate of 10 Hz at room temperature. The MgO film was obtained with an oxygen pressure of 10$^{-3}$ torr and the Pt film was 10$^{-7}$ torr. The target-substrate distance was 60 mm and the focused laser spot size was 1×3 mm$^2$. The Ti/Au layer was obtained by electron beam evaporation.

*Transmission electron microscopy (TEM):* Mn$_3$Ga/PMN-PT samples were thinned by focused Ga ion beam. The surfaces and the crystal structures of Mn$_3$Ga/PMN-PT heterostructures were characterized by a Tecnai G2 F20 TEM setup at 200 kV.

*Electrical & magnetic measurements:* Magnetic measurements were performed in a Quantum Design VersaLab system. The Hall and longitudinal resistivities were measured with the standard four-probe method. The low temperatures needed for Hall resistance measurement were obtained by the Quantum Design VersaLab system. A Keithley 2410 sourcemeter supplied a current of 1 mA. The voltage signals of Hall and longitudinal resistance measurements were detected by a Keithley 2182A nanovolt sourcemeter. As for the measurements of magnetic tunnel junctions, two Keithley 2400 sourcemeters were utilized as a voltage source (50 mV) and a current detector. The gate electric field was applied through another Keithley 2400 sourcemeter.

**Acknowledgements**
Z.L. acknowledges financial support from the National Natural Science Foundation of China (NSFC; grant numbers 51822101, 51861135104, 51771009 & 11704018).

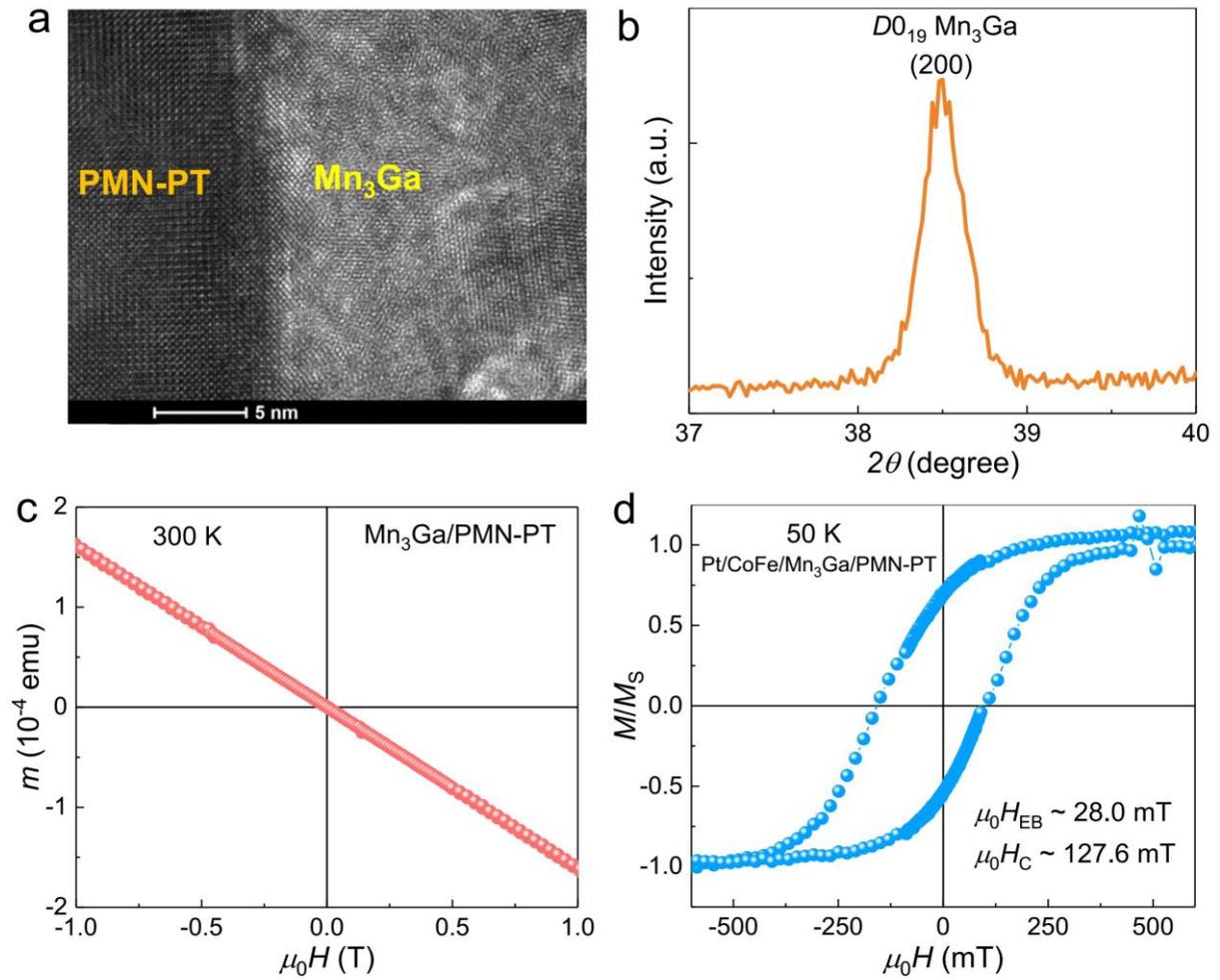

**Figure 1.** a) Cross-section transmission electron microscopy image of a Mn$_3$Ga/PMN-PT heterostructure. The scale bar is 5 nm. b) X-ray diffraction pattern of the Mn$_3$Ga/PMN-PT heterostructure. c) Magnetic moment of the Mn$_3$Ga/PMN-PT structure measured at 300 K. d) Magnetization curve of a 5-nm-thick CoFe layer grown on a 50-nm-thick Mn$_3$Ga film capped by a 2-nm-thick Pt film at 50 K.

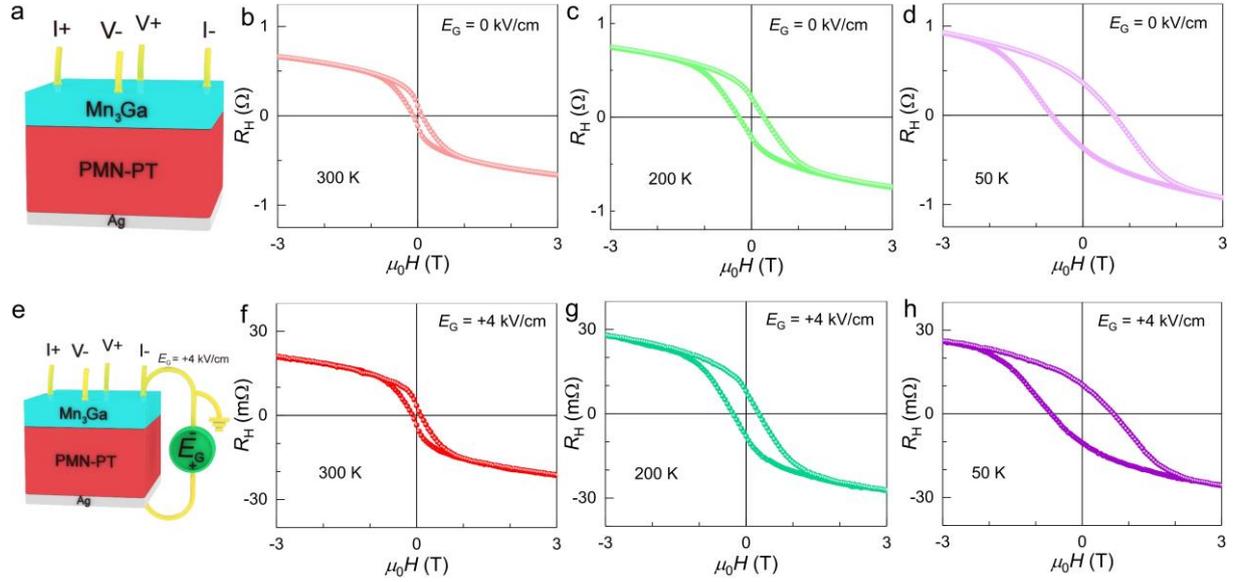

**Figure 2.** a) Schematic of the AHE measurement geometry. Magnetic-field dependent anomalous Hall resistance of the Mn$_3$Ga/PMN-PT heterostructure at 300 K b), 200 K c) and 50 K d). e) Schematic of the electric-field ($E_G$) controlled AHE measurement geometry. Magnetic-field dependent anomalous Hall resistance of the Mn$_3$Ga/PMN-PT heterostructure at $E_G$ = 4 kV/cm at 300 K f), 200 K g) and 50 K h).

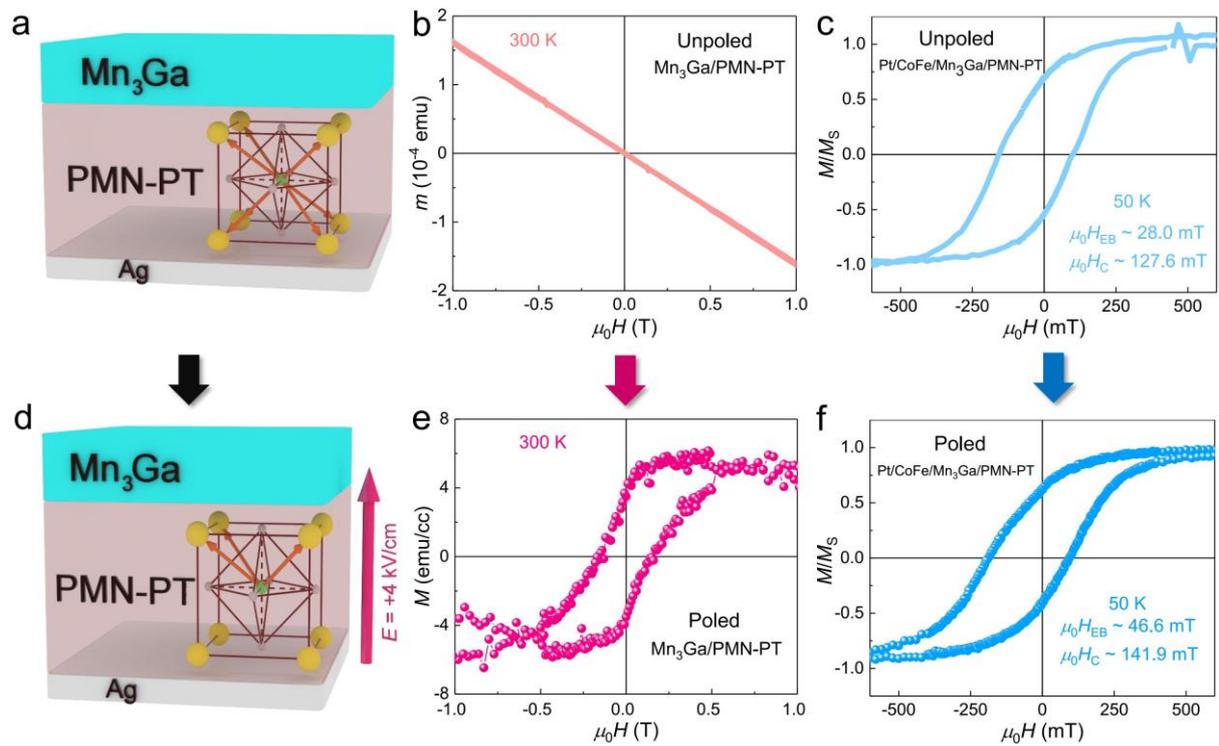

**Figure 3.** a) Schematic of an unpoled Mn$_3$Ga/PMN-PT heterostructure. b) Magnetic-field dependent moment curve of the unpoled Mn$_3$Ga/PMN-PT at 300 K. c) Magnetization versus magnetic field (*M-H*) of a 5-nm-thick Co$_{90}$Fe$_{10}$(CoFe) layer grown on a 50-nm-thick Mn$_3$Ga thin film capped with a 2-nm-thick Pt layer at 50 K. d) Schematic of the Mn$_3$Ga/PMN-PT heterostructure poled by $E_G$ = 4 kV/cm. e) Magnetic-field dependent magnetization curve of the poled Mn$_3$Ga/PMN-PT at 300 K. f) *M-H* loop of the poled Pt/CoFe/Mn$_3$Ga/PMN-PT heterostructure at 50 K.

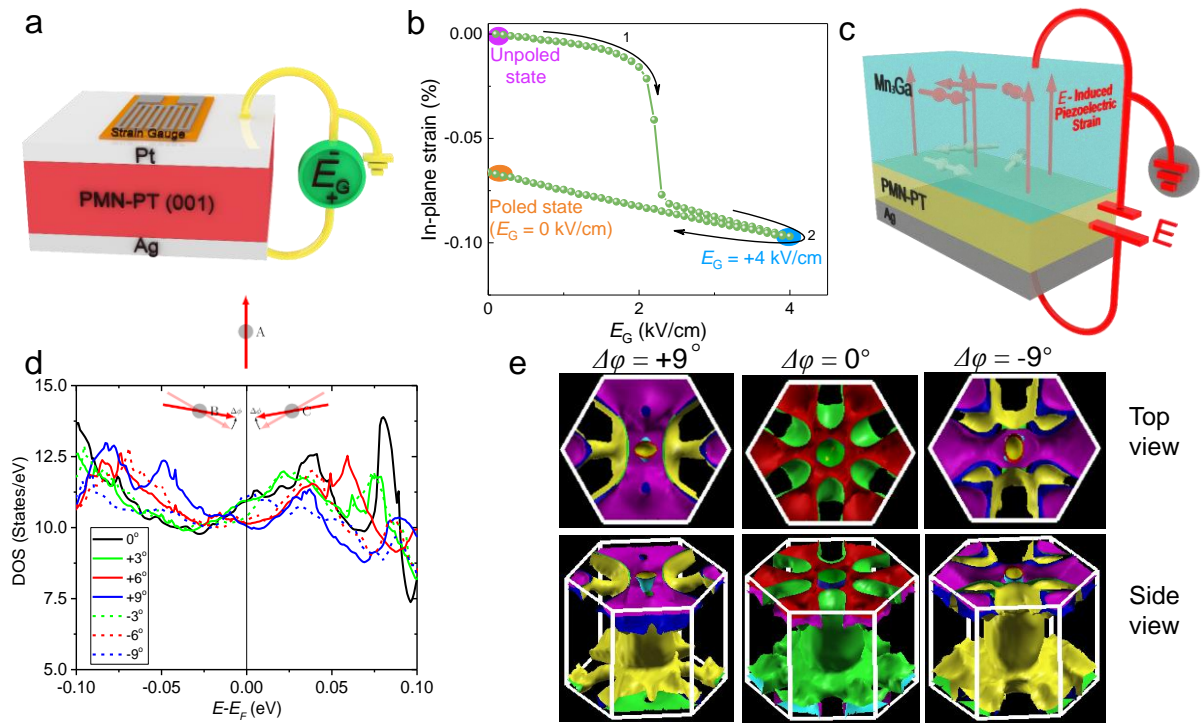

**Figure 4.** a) Schematic of measurement geometry for electric-field ($E_G$) induced in-plane strain of a PMN-PT substrate at 300 K. b) In-plane strain versus $E_G$ of the PMN-PT substrate at room temperature. c) Sketch of the possible transformation of the spin structure of $Mn_3Ga$ thin film from unpoled state (white) to poled state (red). d) Density of states for the noncollinear spin structure of $Mn_3Ga$ with different spin rotation angles. Inset: Schematic of an upward spin rotation $\Delta\varphi$ of spin B and spin C relative to the fixed spin A under biaxial in-plane compressive piezoelectric strain for a triangular spin structure consisting of three spins A, B and C. e) Top and side views of three-dimensional Fermi surfaces of the $Mn_3Ga$ for the 120° triangular spin structure and the noncollinear spin structures rotated by positive and negative 9°. The complete Fermi surface is derived from multiple energy bands and different portions of the Fermi surface related to different energy bands are color-coded.

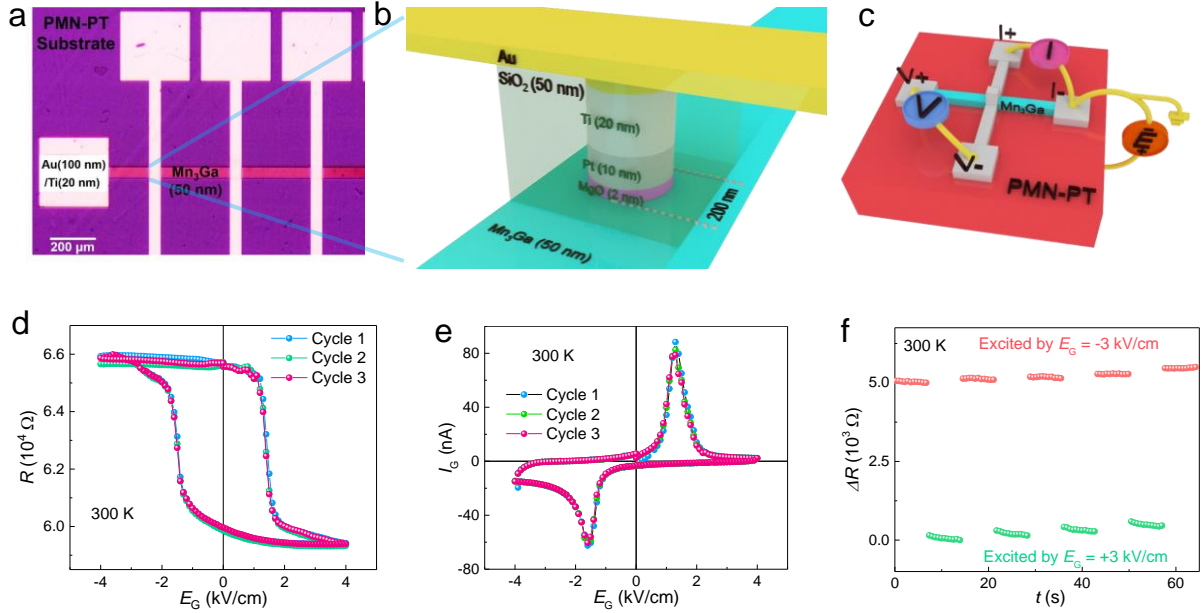

**Figure 5.** a) Optical microscope image of MTJs with Ti/Au top electrodes. The scale bar is 200 μm. b) Schematic of the MTJ stacking structure. c) Schematic for the measurement geometry of electric-field-controlled MTJ. d) $E_G$-dependent resistance-area product ($RA$) curves of the MTJ during three continuous measurements. e) $E_G$-dependent gate current ($I_G$) of the PMN-PT substrate. f) The high-resistance and the low-resistance states realized by electric pulse of $E_G$ = -3 and +3 kV/cm, respectively, at 300 K.